\documentclass[aps,pra,twocolumn,superscriptaddress]{revtex4}
\usepackage{amsmath,amsthm}
\usepackage{epsf}
\usepackage{amssymb}
\usepackage{graphicx}
\usepackage{color}      

\newcommand{\ket}[1]{| #1 \rangle}

\newcommand{\rb}[1]{\left( #1 \right)}

\newcommand{\ex}[1]{\langle #1 \rangle}

\newcommand{\beq}{\begin{eqnarray}}
\newcommand{\eeq}{\end{eqnarray}}

\begin{document}
\title{Entanglement and Entropy in a Spin-Boson Quantum Phase Transition.}
\author{N.~Lambert}
\affiliation{The University of Manchester, P.O. Box 88,
Manchester, M60 1QD, U.K.}
\author{C.~Emary}
\affiliation{Instituut-Lorentz, Universiteit Leiden, P.O. Box 9506,
2300 RA Leiden, The Netherlands}
\author{T.~Brandes}
\affiliation{The University of Manchester, P.O. Box 88,
Manchester, M60 1QD, U.K.}
\date{\today}
\bibliographystyle{apsrev}
\begin{abstract}
We investigate the entanglement properties of an ensemble of atoms
interacting with a single bosonic field mode via the Dicke
(superradiance) Hamiltonian.  The model exhibits a quantum phase
transition and a well-understood thermodynamic limit, allowing the
identification of both quantum and semi-classical many-body
features in the behaviour of the entanglement. We consider
the entanglement between the atoms and the field, an investigation 
initiated in \cite{Lambert04}.
 In the thermodynamic
limit, we give exact results for all entanglement partitions and
observe a logarithmic divergence of the atom-field entanglement,
and discontinuities in the average linear entropy.
\end{abstract}
\maketitle

\section{Introduction}

Understanding entanglement -- the quantum correlations impossible to
mimic with local classical theories --
is a fundamental goal of quantum information science. Similarly,
understanding complex modes of behaviour, such as quantum phase
transitions~\cite{Sachdev99} and quantum
chaos~\cite{Gutzwiller90}, has become an important part of quantum
many-body theory. Since large correlations and collective behaviour
are an intrinsic part of critical systems, concepts and formalisms
used to describe entanglement are now being employed to reveal the
truly quantum nature of certain aspects of criticality.

Investigations into the entanglement between interacting spin-1/2
systems on a one dimensional chain have revealed so-called
`critical entanglement', in which an entanglement measure of the
ground state exhibits universality, or scaling behaviour, around
the critical point \cite{Osborne02}. In particular, for the infinite XY spin chains,
(and their Ising variants) it has been shown that entanglement
between nearest and next-nearest neighbours reaches a maximum {\em
near}, but not at, the critical point \cite{Osterloh02,Osborne02}.
Furthermore, Osterloh {\it et al}. \cite{Osterloh02} have observed
scaling behaviour of the entanglement, showing that the derivative
of the concurrence diverges logarithmically near the critical
point. They also found a logarithmic divergence of the derivative
as a function of system size.

Latorre, Vidal, and co-workers \cite{Vidal02, Latorre03} took a
different approach and investigated the entanglement, via the von
Neumann entropy, between a block of $L$ spins and the rest of the
chain in $XY$ and Heisenberg spin-chains.  They found a
logarithmic scaling of the entropy with $L$; this time with a
pre-factor corresponding to the `central charge' of a $1+1$
continuum quantum field theory of the same universality class.  In
effect, they found the same area law associated with the
geometric entropy first studied by Srednicki \cite{Srednicki93}.
In an effort to understand the nature of the scaling of
entanglement, Or\'{u}s {\it et al.} \cite{Orus03} illustrate that
the scaling of the entanglement at the critical point determines
whether or not one could efficiently simulate the quantum system
at this point on a classical computer.

Going beyond 1-dimensional spin chains, the authors of
\cite{Vidal03} studied a highly connected simplex, where each spin
interacts equally with all other spins, and the lattice spacing no
longer plays an important role. Importantly, because of the
symmetry, they find a maximum in the pairwise concurrence at the
critical point, and determine scaling exponents for the behaviour
of the concurrence with system size.

In this paper, we continue the investigation, begun in \cite{Lambert04}, of the entanglement properties of the
single-mode Dicke Hamiltonian, which describes an ensemble of $N$
two-level atoms coupled to a single-mode bosonic field. This model
exhibits a `superradiant' quantum phase transition (QPT)
in which the ground state
undergoes a dramatic change in character.  We consider
several aspects of the ground-state entanglement in this model and
observe how they are affected by the QPT. We investigate
entanglement between the atomic ensemble and the field mode via
the von Neumann \cite{Wehrl78,Schumacher93} and linear entropies of this
bipartite decomposition.  We also calculate the average linear entropy of all the subsystems, which
corresponds to a multipartite measure introduced by Meyer and Wallach \cite{Meyer01}.
In the thermodynamic limit, the model is exactly soluble across the whole
coupling range, and we give exact results for these quantifiers of
the entanglement. For finite $N$ we use perturbative and numerical
methods.

In summary, we find the atom-field entropy diverges at the phase
transition alongside the traditional correlation length, with
corresponding critical exponents, and may be fruitfully described
by an effective `entanglement temperature'.  
As has been discussed previously \cite{Emary202, Emary02}, the QPT
is foreshadowed at finite $N$ by various `precursors', and in
particular, a transition from integrable to Quantum Chaotic
behaviour near the critical point. This transition is
characterised by a change in the energy level statistics, and can be
correlated with the change in the phase-space of a classical
Hamiltonian corresponding to the Dicke model. The phase transition in
the quantum model maps to a supercritical pitchfork bifurcation in
the classical model, and such bifurcations have recently been
related to entanglement characteristics
\cite{Hines03,Schneider02}.  In addition, Fujisaki {\it et al.}
have shown that the appearance and strength of chaos can be linked
to the production of entanglement \cite{Fujisaki03}. Further work
is required in clarifying the relation between entanglement in
quantum systems and chaos in the corresponding classical model.
However, there is a conceptual connection between the divergence
of trajectories in classical chaos and the delocalization of the
quantum ground state, which is, in general, indicative of
entanglement.

The model considered here is of wider interest still, given that the
interaction of a charge or spin systems with a single bosonic mode
is viewed as a mechanism for generation of entanglement in many
different situations such as quantum cavity QED, quantum dots
\cite{Vorrath02,Wang02}, and ion traps.  In addition, many
suggestions have been made to use the environment, or bosonic
cavities, to share or mediate
entanglement \cite{Wang03,Plenio02,Vorrath02,Lambert03}.  In particular,
Reslen et al \cite{Reslen04} have shown that there is a direct equivalence between
the single mode Dicke Hamiltonian and the infinitely-coordinated XY model.

This paper has the following structure.
In section \ref{secDH} we
reintroduce the Dicke Hamiltonian, and describe the quantum phase
transition. In section \ref{secAF} we consider the atom-field
entanglement by recalling the finite numerical
and exact thermodynamic limit results for the von-Neumann entropy considered in \cite{Lambert04}, and, as mentioned, extending
the discussion with a calculation of the linear entropy, participation ratio, and the average linear entropy.  
We omit discussion of the pairwise entanglement covered in \cite{Lambert04}. We conclude with discussions
in section \ref{secDISC}.

\section{The Dicke Model}\label{secDH}
Generically, the Dicke Hamiltonian (DH) describes the dipole
interaction between $N$ atoms and $n$ bosonic field modes.  Here
we shall only consider the single mode case with $n=1$. A standard
approach to such quantum-optics Hamiltonians is to make the
rotating wave approximation (RWA), rendering the model integrable.
We do not make the RWA here, allowing the model to describe both
weak and strong coupling regimes.

\subsection{The Hamiltonian}
The single-mode Dicke Hamiltonian is
\beq H &=&
   \omega_0 \sum_{i=1}^N s_z^{(i)}
    + \omega  a^\dagger a
    + \sum_{i=1}^N
        \frac{\lambda}{\sqrt{N}} \rb{a^\dagger + a}
        \rb{s^{(i)}_+ + s^{(i)}_-}
  \nonumber \\
  &=&
  \omega_0 J_z + \omega a^\dagger a
  + \frac{\lambda}{\sqrt{2j}} \rb{a^\dagger + a}\rb{J_+ + J_-},
\label{DHam1} \eeq where $J_z=\sum_{i=1}^N s_z^{i}$,
$J_{\pm}=\sum_{i=1}^N s_{\pm}^{i}$ are collective angular momentum
operators for a pseudo-spin of length $j=N/2$.  These operators
obey the usual angular momentum commutation relations,
$[J_z,J_{\pm}]=\pm J_{\pm}$ and $[J_+,J_-]= 2J_z$. The frequency
$\omega_0$ describes the atomic level splitting, $\omega$ is the
field frequency, and $\lambda$ the atom-field coupling
strength.

There exists a conserved parity operator
\beq
  \Pi = e^{i\pi(a^{\dagger}a + J_z + j)},
\eeq which commutes with the Hamiltonian.  For finite $N$, the
ground state has positive parity.  The DH undergoes a QPT at a
critical value of the atom-field coupling $\lambda_c =
\sqrt{\omega \omega_0}/2$ which breaks this symmetry.

At finite $N$, we perform numerical diagonalisations
using a basis $\ket{n} \otimes \ket{j,m}$, where $\ket{n}$ are
Fock states of the field, and $\ket{j,m}$ are the so-called Dicke
states -- eigenstates of $\mathbf{J}^2$ and $J_z$.  
We make use of the parity symmetry to simplify these numerics.

\subsection{Thermodynamic Limit}
The DH undergoes a QPT in the thermodynamic limit ($j\rightarrow
\infty\Leftrightarrow N\rightarrow \infty$, notation which we will
use interchangeably) at a critical coupling of $\lambda_c =
\sqrt{\omega \omega_0}/2$. Below $\lambda_c$ the system is in its
normal phase in which the ground state is largely unexcited. Above
$\lambda_c$, the superradiant phase, the ground-state possesses a
macroscopic excitation.

As illustrated in Ref. \cite{Emary202}, exact solutions may be
obtained for both phases in the thermodynamic limit by employing
a Holstein-Primakoff transformation of the angular momentum
algebra. In this section, we briefly summarise this analysis,
highlighting those features such as are required here.

The Holstein-Primakoff mapping expresses the angular momentum in
terms of a single boson mode,
\beq
  J_+ = b^{\dagger}\sqrt{2j - b^{\dagger}b},\quad J_- =\sqrt{2j -
  b^{\dagger}b} b,\quad J_z = b^{\dagger}b - j,
\eeq
with $[b,b^{\dagger}]=1$. In
this representation, the DH transforms into a two mode bosonic
problem.

\subsubsection{Normal Phase}
The normal phase is found by simply taking $j \rightarrow \infty$ in the bozonised hamiltonian, which produces a linear two mode hamiltonian.  This, as described in \cite{Lambert04}, can
be diagonalised with a Boglioubov transformation.

To calculate the atom-field entanglement of the ground state,
we require the reduced density matrix (RDM) of the atoms in the ground state. 
Summarising our steps in \cite{Lambert04}, the
ground-state wave function is a product of two Gaussians,
$\Psi(q_1,q_2) = G_+(q_1) G_-(q_2)$ described by the co-ordinates
corresponding to the bosonic operators of the diagonalised Hamiltonian. Inverting the Boglioubov
coordinate rotations gives us the wave function in terms of the
coordinates $(x,y)$ corresponding to the physical field ($x$) and
atom $(y)$ modes.  To obtain the RDM of the atomic system, we integrate over the $x$
coordinate.  We write the resulting RDM in terms of a rescaled
$y$ coordinate $y\rightarrow y/\kappa$ (writing $c= \cos \gamma^{(1)}$, and $s=\sin
\gamma^{(1)}$, $\tan (2\gamma^{(1)}) = \frac{4\lambda \sqrt{\omega \omega_0}}{(\omega_0^2 - \omega^2)}$),
\beq
  \label{rdm} \rho_G(y,y') &=&
  \rb{\frac{\epsilon_{+}^{(1)}\epsilon_{-}^{(1)}}{\pi(\epsilon_-
  c^2 + \epsilon_+ s^2)}}^{\frac{1}{2}}
  \nonumber \\
  &\times& \exp \left( \frac{2\epsilon_- \epsilon_+ +
  D}{4\kappa^2 (\epsilon_- c^2 + \epsilon_+ s^2)} (y^2 + y'^2) \right.
  \nonumber  \\
  &+& \left. \frac{D}{2\kappa^2(\epsilon_- c^2 + \epsilon_+ s^2)}y
  y'\right),
\eeq
where $D=(\epsilon_- - \epsilon_+)^2c^2s^2$, and the excitation energies in this normal phase are
\beq
\epsilon_{\pm}^2=\frac{1}{2}\left(\omega_0^2 + \omega^2 \pm \sqrt{\left(\omega_0^2 - \omega^2\right)^2 + 16\lambda^2\omega \omega_0} \right).
\eeq
We did not perform this rescaling in \cite{Lambert04}.  As this rescaling is effected by a unitary transformation on the atomic
system alone, it will not affect the atom-field entanglement. It does,
however, aid in the interpretation of our results, as we show later.
Note that the RDM for the field mode is the same as above,
except with $c$ and $s$ interchanged.

\subsubsection{Super-Radiant Phase}
In the following
section, we describe in more detail the calculations and properties of this phase
which were not covered in \cite{Lambert04}.  In the superradiant (SR) phase ($\lambda>\lambda_c$), both atom
and field degrees of freedom acquire macroscopic mean-fields.  We
incorporate these mean-fields by displacing the two oscillator modes
\beq
  a^{\dagger} &=& c^{\dagger} \pm \sqrt{\alpha},
  \quad b^{\dagger} = d^{\dagger} \mp \sqrt{\beta},
  \label{disp}
\eeq
where $\alpha$, $\beta$ are of order $j$.  That there are two
choices of sign here is significant, as the two choices lead to
two different Hamiltonians with degenerate solutions -- an indication that the parity of the system
has been broken in this phase.

By inserting one of the above displacements into the Holstein-Primakoff bozonised Hamiltonian, and setting terms with overall powers of $j$ in the
denominator to zero, we obtain an exactly soluble Hamiltonian.
Diagonalization requires a specific choice for the displacements
$\sqrt{\alpha}=\frac{2\lambda}{\omega}\sqrt{\frac{j}{2}(1-\mu)}$,
$\sqrt{\beta}=\sqrt{j(1-\mu)}$, with $\mu= \lambda_c^2 / \lambda^2$,
and a rotation of the coordinates \beq
  Q_1 = X \cos \gamma^{(2)} - Y \sin \gamma^{(2)},
  \nonumber \\
  Q_2 = X \sin \gamma^{(2)} + Y \cos \gamma^{(2)},
\eeq
with angle of rotation given by
\beq
  \tan (2\gamma^{(2)}) = \frac{2\omega \omega_0
  \mu^2}{\omega_0^2 - \mu^2 \omega^2}.
\eeq
The excitation energies of the SR phase are
\beq
  \epsilon_{\pm}^2=\frac{1}{2}\rb{{\frac{\omega_0^2}{\mu^2} +
  \omega^2 \pm \sqrt{\rb{\frac{\omega_0^2}{\mu^2} - \omega^2}^2 +
  4\omega^2 \omega_0^2}}}.
\eeq
and $\epsilon_-$ is real only for
$\lambda \geq \lambda_c$.  The effective Hamiltonians derived with either choice of sign in
Eq. (\ref{disp}) do not commute with the parity
operator $\Pi$, and thus we see that this symmetry is broken in the SR phase.

As before, the ground state of the diagonalised Hamiltonian is the
product of two Gaussians in $Q_1$ and $Q_2$.  To obtain the wave
function in terms of the original atomic and field co-ordinates,
we must not only perform the rotation $Q_1, Q_2 \rightarrow X,Y$
but also take into account the relationship between the displaced
and re-scaled coordinates $X,Y$ and the original atom-field
co-ordinates $x,y$, \beq
  X=x \mp \sqrt{\frac{2\alpha}{\omega}},\quad
  Y=\sqrt{\frac{\omega_0}{\widetilde{\omega}}}y \pm
  \sqrt{\frac{2\beta}{\widetilde{\omega}}},
\eeq
where $\widetilde{\omega}=\frac{\omega_0}{2\mu}(1+\mu)$.
In the displaced ($X,Y$) frame the wave functions have the
same form as in the normal phase, but with different parameters and
coefficients. In the original frame ($x,y$) the wave functions
are again the same but displaced from the origin.

\subsubsection{Two Lobes}

The two possible displacements lead to two Hamiltonians with ground state wave functions in the $x,y$ representation displaced from the origin in  different directions.  For large but finite $N$, the  ground state in the SR phase is a single two-lobed wave function, 
\begin{widetext}
\beq\label{SR2lobes}
  \Psi_G^\mathrm{SR}(x,y)|_{N\gg 1} &\approx& \frac{1}{\sqrt{2}}\left( \sum_{\pm}
  G_-\left[(x\pm \sqrt{\frac{2\alpha}{\omega}})c-
  \sqrt{\frac{\omega_0}{\widetilde{\omega}}}(y\pm\sqrt{\frac{2\beta}{\omega_0}})s\right]
   G_+\left[(x\pm\sqrt{\frac{2\alpha}{\omega}})s +
  \sqrt{\frac{\omega_0}{\widetilde{\omega}}}(y\pm\sqrt{\frac{2\beta}{\omega_0}})c\right]\right)
\eeq
\end{widetext}
where $G_{\pm}$ are the normalised Gaussians. This state has positive parity and can be used for comparison with our numerical results at finite $N$: since the displacements are of order $\sqrt{N}$, for $N\gg 1$ the lobes have exponentially small overlap whence the reduced density matrix becomes 
\begin{widetext}
\beq \label{2lobe}
  \rho^{SR}(y,y')|_{N\gg 1} &\approx& \frac{1}{2}\sum_\pm
  \rho_G\left[\sqrt{\frac{\omega_0}{\widetilde{\omega}}}\left(y\pm\sqrt{\frac{2\beta}{\omega_0}}\right),
  \sqrt{\frac{\omega_0}{\widetilde{\omega}}}\left(y'\pm\sqrt{\frac{2\beta}{\omega_0}}\right)\right],
\eeq
\end{widetext}
where $\rho_G$ is given by \ Eq.~(\ref{rdm}).

\section{Atom-field Entanglement}\label{secAF}
The RDMs of the atoms, \ Eq.~(\ref{rdm}){} and \ Eq.~(\ref{2lobe}){}, are derived from the  pure ground states in the normal and SR phase, respectively. The atom-field entanglement is therefore determined by the von Neumann
entropy \beq
  S= - \mathrm{Tr}(\rho \log_2 \rho)
\eeq
with $\rho$ the RDM of the atoms (an identical result is obtained with the field RDM).
We present first 
numerical results (already discussed in \cite{Lambert04}) and a new perturbative result for finite $N$, and then recap our exact solutions in the
thermodynamic limit.

\begin{figure}[]
\includegraphics[width=0.5\textwidth]{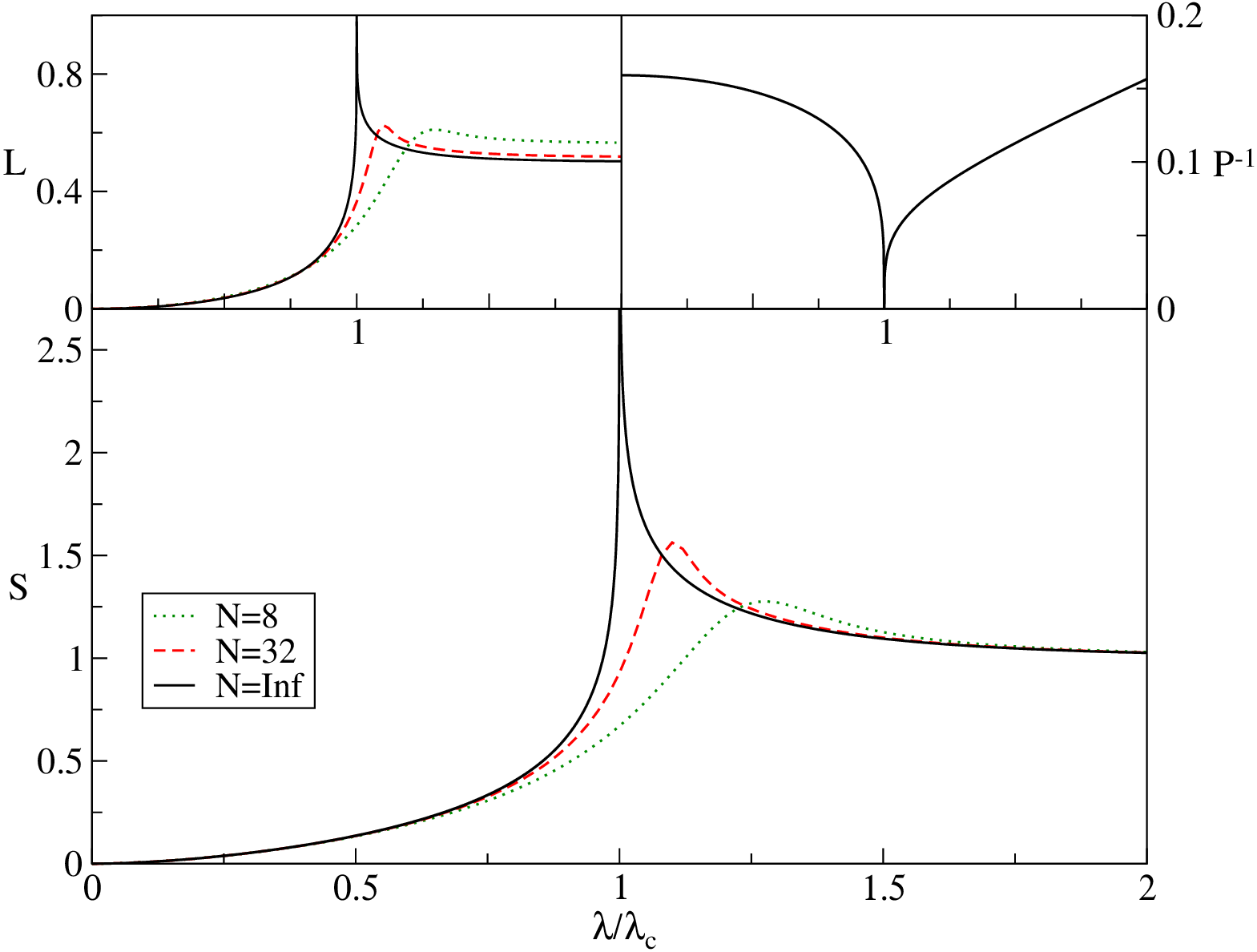}
\caption{The von Neumann entropy $S$ between the atomic and
bosonic modes for a range of system sizes, $N=8$, $N=32$, and
$N\rightarrow\infty$. The divergence at $\lambda/\lambda_c = 1$ can be
clearly seen, as well as the strong coupling $S=1$ limit. The
smaller graphs depict the linear entropy $S_{\text{lin}}$ between
the atomic and bosonic modes Eq. (\ref{slin}), and the inverse participation ratio
$P^{-1}$ Eq. (\ref{part}) of the ground state wave-function.  The linear entropy
exhibits the similar behaviour to the von Neumann entropy, except
the asymptotic limit changes with $N$ because of the normalisation. 
The participation ratio drops to $0$ at the critical point,
indicating a massive delocalization of the ground state.}
\label{thermvne}
\end{figure}

We diagonalise the DH in the Fock-Dicke basis, and obtain the RDM of the atoms.
This is diagonalised and the von Neumann entropy is obtained from
\begin{equation}
  S(\rho) = -\sum_k p_k \log_2 p_k.
\end{equation}
where $p_k$ are the eigenvalues of the RDM \cite{Wehrl78}.  In
Fig. \ref{thermvne} we plot the results of these numerical
calculations.

\subsection{Perturbative Results}
>From Rayleigh-Schr\"odinger perturbation theory, we find an
$N$-independent result for the von Neumann entropy for low
coupling (with $\sigma =\lambda / (\omega + \omega_0)$)
\begin{equation}
S = - \frac{1}{1+ \sigma^2} \log_2\rb{\frac{1}{1+ \sigma^2}} -
\frac{\sigma^2}{1+ \sigma^2}\log_2\rb{\frac{\sigma^2}{1+
\sigma^2}}.
\end{equation}
This matches the numerical data well for $\lambda/\lambda_c
\lesssim 0.4$. Similarly, following Refs.~\cite{Emary01, Frasca03}
for $\lambda \rightarrow \infty$, we can identify the strong
coupling limit ground state as $\ket{\Psi_{GS}} =
\frac{1}{\sqrt{2}} \rb {\ket{\frac{\sqrt{2j}\lambda}{\omega},
-j_x} + \ket{-\frac{\sqrt{2j}\lambda}{\omega}, j_x}}$, where
$\ket{\pm \frac{\sqrt{2j}\lambda}{\omega}, \mp j_x}$ is a product
of a coherent state for the field and an eigenstate of $J_x$ for
the atoms.  As this is effectively a maximally entangled state of
two two-level systems, $S\rightarrow 1$ as $\lambda \rightarrow
\infty$.

\subsection{Thermodynamic limit}
For $N\to \infty$, the excitation energy
diverges as $\epsilon_- \propto |\lambda_c - \lambda|^{z \nu}$ and
the characteristic length diverges as $l_- =
\frac{1}{\sqrt{\epsilon_-}} \propto |\lambda - \lambda_c|^{-\nu}$
with the exponents $z=2$ and $\nu= 1/4$ \cite{Emary202}.
We now use the thermodynamic limit RDMs found above to obtain an exact
analytical expression for the entropy $S$ as $N\rightarrow \infty$.
This calculation proceeds via comparison with the density matrix of a single harmonic
oscillator of mass $m$, frequency $\Omega$ at temperature $T$
\cite{Feynman98,Lambert04} with that of our reduced atomic system in Eq.(\ref{rdm}).  We find,
\beq \label{effT}
  \cosh \beta \Omega &=& \rb{1 + \frac{2\epsilon_-
  \epsilon_+}{(\epsilon_- - \epsilon_+)^2 c^2 s^2}},
   \\
  m \Omega &=& \sqrt{\rb{1+\frac{2\epsilon_-\epsilon_+}{(\epsilon_-
  - \epsilon+)^2 c^2 s^2}}^2 - 1} \nonumber \\
  &\times& \rb{\frac{(\epsilon_- - \epsilon_+)^2 c^2
  s^2}{2\kappa^2(\epsilon_-c^2 + \epsilon_+ s^2)}}
  \label{fT}
\eeq where $\beta = 1/k_B T$.  We have two equations linking the
four parameters of the atomic RDM $\omega$, $\omega_0$, $\lambda$,
$\kappa$ (where $\kappa$ is the squeezing parameter we introduced
in Eq.(\ref{rdm})) and the three effective parameters of the
thermal oscillator $\beta$, $\Omega$, $m$. By setting one energy
scale of the original system such that $\omega_0=1$, and that of
the thermal oscillator such that $m=1$, $\Omega=\omega$, we can
uniquely define the correspondence between the two systems.

The squeezing parameter $\kappa$ introduced into the RDM in Eq.
(\ref{rdm}) compensates for the one-mode squeezing that the atomic
ensemble undergoes as a function of  $\lambda$ \cite{Emary02},
allowing us to keep the frequency of the thermal oscillator
constant.  With this relation between the parameters of the two
RDMs, the effective temperature becomes the parameter describing
the degree of mixing in the RDM. In other words, the interaction
of the field with the atomic ensemble is such that, from the point
of the atoms alone, it is as if they were at a finite temperature,
with the temperature given by Eqs (\ref{fT}). The determination of
this temperature is not unique, since there are more free
parameters in Eqs (\ref{fT}) than constraints, but the choice made
here is physically appealing, with the frequency of the thermal
oscillator constant and the temperature varying with
$\lambda$.

\begin{figure}[]
\includegraphics[width=0.5\textwidth]{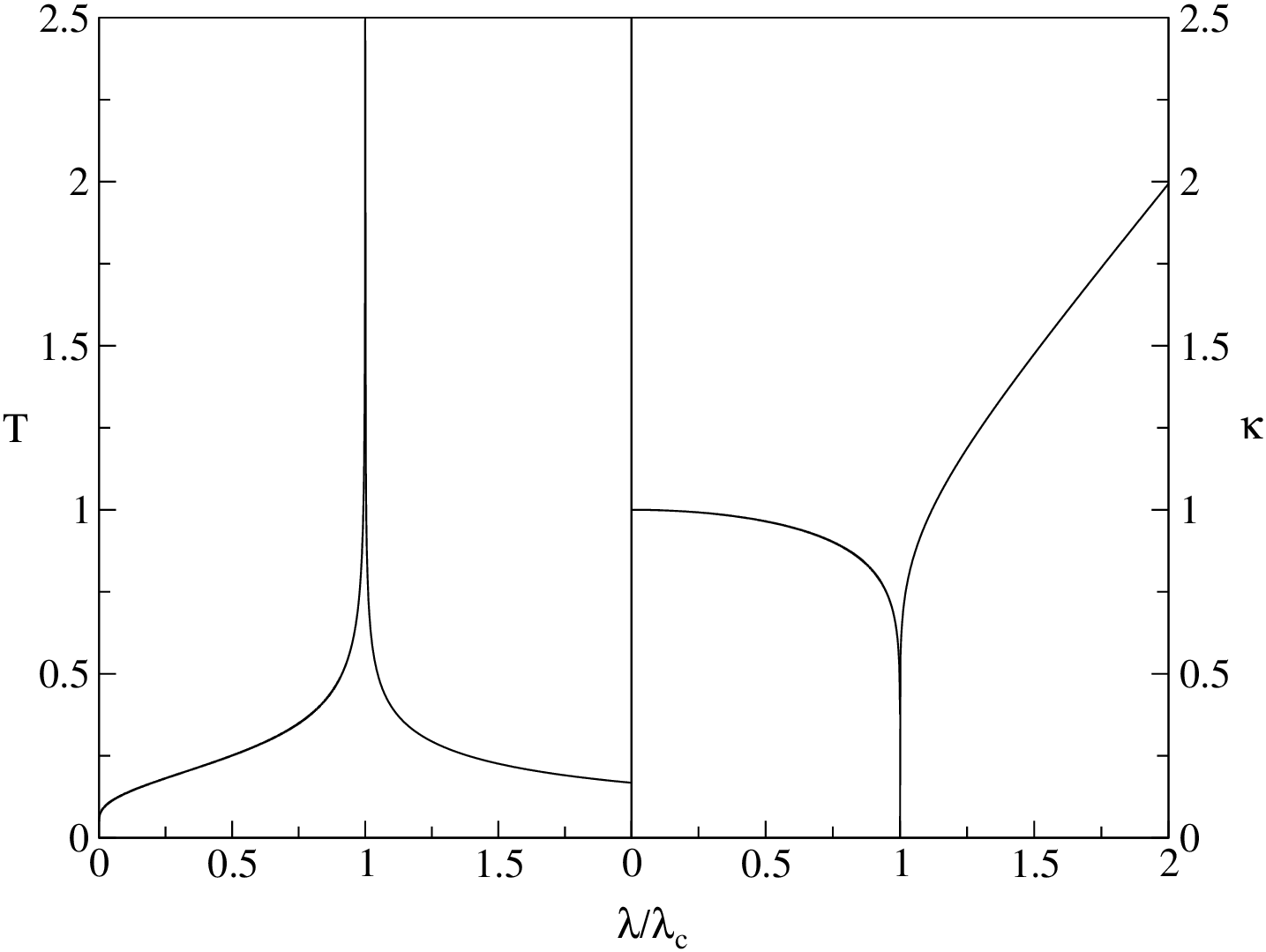}
\caption{Effective entanglement temperature $T$ vs.
$\lambda/\lambda_c$ for $N\rightarrow \infty$,
$\omega=\omega_0=1$, from Eq. (\ref{effT}) with $\beta = 1/T$. The
divergence at the critical point causes the simultaneous
divergence of the von Neumann entropy, signalling `maximal
mixing'.  The second graph illustrates the behaviour of the
squeezing parameter $\kappa$ vs $\lambda/\lambda_c$ from Eq.
(\ref{fT}), which tends to $0$ at the critical point.}
\label{therm}
\end{figure}

The behaviour of the temperature $T$ with $\lambda$ is shown in
Fig. \ref{therm}, and the divergence at the critical point is
immediately obvious.  We also plot the squeezing parameter
$\kappa$, which vanishes at $\lambda_c$ in accordance with the
delocalization of the system here.

The entropy of a harmonic oscillator at finite
temperature is a standard result from statistical physics
\cite{Feynman98} (setting $\hbar=k_B=1$),
\begin{equation}\label{entropy}
  S= \rb{\frac{\Omega}{2T} \coth (\frac{\Omega}{2T}) -
  \ln{(2\sinh(\frac{\Omega}{2T}))}}/\ln (2).
\end{equation}
Note that this is independent of $\kappa$, and thus the above
discussion does not affect the result for $S$.
Solving Eqs (\ref{fT}) for the effective
parameters, we obtain the von Neumann entropy of the atom-field
system in the normal phase, which is plotted in
Fig. \ref{thermvne}. We clearly see a divergence at $\lambda_c$.

Moving into the SR phase, if we calculate the entropy of a single
displaced lobe, exactly the same calculation as in the normal phase
applies, except with the SR parameters instead of the normal phase
ones. Around the critical point, the entropy diverges and then
falls to zero for large coupling 
(not shown here). This is the correct scenario for $N\to \infty$, where 
parity symmetry is broken and the system sits in either of the displaced lobes.

The more interesting case occurs for large but finite $N$ where our 
numerical results indicate that for large coupling, the entropy $S$ does not tend to
zero, but rather to a finite value. This can 
be easily understood by calculating the entropy
of the positive-parity two-lobed SR RDM of Eq. (\ref{2lobe}), 
rather than the broken-parity single-lobe wave function as above.
We recall that  the two-lobe RDM for $N\gg 1$ formally turns out as a mixture of the density
matrices representing each lobe, cf. \ Eq.~(\ref{2lobe}){}. 
A standard result~\cite{Wehrl78}
is that for a density matrix $\rho= \lambda_1 \rho_1 + \lambda_2
\rho_2$, $\lambda_{1,2}\ge 0$, the concave nature of the von
Neumann entropy allows us to write the following inequality,
\begin{equation}
  S(\rho) \geq \lambda_1 S(\rho_1) + \lambda_2 S(\rho_2).
  \label{sb1}
\end{equation}
The entropy of a mixture  of density matrices is also bounded from above by
\begin{equation}
  S(\rho) \leq \lambda_1 S(\rho_1) + \lambda_2 S(\rho_2) - \lambda_1
  \log \lambda_1 - \lambda_2 \log \lambda_2.
  \label{sb2}
\end{equation}
The final two terms are known as the mixing entropy \cite{Wehrl78},
and in the case that the ranges of the $\rho_1$ and $\rho_2$ are pairwise orthogonal,
this upper bound becomes an equality. Returning to our positive-parity SR RDM, the entropy of each of
the two lobes is identical $S(\rho_1) = S(\rho_2)$ and they are
weighted in an equal superposition $\lambda_1=\lambda_2=1/2$.
Furthermore, the two lobes are orthogonal, and thus from Eqs
(\ref{sb1},\ref{sb2}) we have
\begin{equation}
  S(\rho) = S(\rho_{1}) + 1.
\end{equation}
We emphasize that the SR phase entropy plotted in Fig.(\ref{thermvne}) is a consistent entanglement measure based on the underlying pure ground state \ Eq.~(\ref{SR2lobes}){}, yielding the correct large-$N$ behaviour for strong couplings $\lambda$.

From Eq. (\ref{entropy}) we see the entropy depends only on the
ratio $\Omega/T$.  In the limit $\lambda \rightarrow \lambda_c$,
we have $\Omega/T \sim \sqrt{\epsilon_-}$. Since as $\lambda
\rightarrow \lambda_c$, $\epsilon_- \rightarrow 0$, we see that
the effective temperature diverges $T\rightarrow \infty$ and so
does the entropy, $S \rightarrow \infty$.
As discussed in \cite{Lambert04}, in the neighbourhood of the critical point, we have
\begin{eqnarray}
  S_{\lambda \rightarrow \lambda_c} &=& [1 - \frac{1}{4} \ln (\frac{64
  \lambda_c^3 \omega^4}{16 \lambda_c^4 + \omega^4}) - \frac{1}{4}
  \ln |\lambda_c - \lambda|]/\ln(2)\nonumber \\ &=& - \frac{1}{4}
  \log_2|\lambda_c - \lambda| + \text{const}.
\end{eqnarray}
The prefactor to the logarithmic divergence is identical to the
exponent characterising the divergence of the length scale $\nu =
1/4$. Thus we see that, as adjudged by the atom-field entropy, the
system is critically entangled.



\subsection{Linear Entropy, Participation Ratio, and the Average Linear Entropy}
An alternative measure of entanglement is the linear entropy,
given by \beq \label{line} L=\eta[1-\text{Tr}(\rho^2)], \eeq where
$\rho$ is the reduced density matrix of one part of our bipartite
system, and $\eta$ is the normalisation $\eta=1+(1/N)$ which gives the correct $0\leq L\leq 1$ behaviour \cite{sco04}.  While it is a valid monotonic entanglement measure, it
lacks some of the full physical interpretation provided by the von
Neumann entropy \cite{Schumacher93,Bennett96}. Again, we calculate
explicit analytical expressions in the thermodynamic limit by
employing our co-ordinate space ground state (noting for $D\rightarrow \infty$, $\eta \rightarrow 1$), \beq
  \text{Tr}(\rho^2)&=&\int dx dx' \rho(x,x') \rho(x',x) \\
  &=& \int dx dx' dy dy' \psi(x,y)\psi(x',y)\psi(x,y')\psi(x',y').
  \nonumber
\eeq
In the normal phase
\beq
  L&=&1-\rb{\frac{\epsilon_- \epsilon_+}{(\epsilon_- c^2 + \epsilon_+ s^2)}}
  [(\epsilon_- s^2 + \epsilon_+ c^2)^2 \\
  &-& \frac{(\epsilon_- s^2 + \epsilon_+ c^2)(\epsilon_- -
  \epsilon_+)^2 c^2 s^2}{(\epsilon_-c^2 + \epsilon_+
  s^2)}]^{-1/2},\nonumber
\eeq
On resonance when $\omega=\omega_0=1, c=s=1/\sqrt{2}$
this simplifies  to
\beq\label{slin}
  L = 1 -
  \frac{2\sqrt{\epsilon_- \epsilon_+}}{(\epsilon_- + \epsilon_+)},
\eeq
which is zero at zero coupling, and unity at the critical
point. In the SR phase we recall the ground state (for large but finite $N$) is a
superposition of two lobes, and the RDM is a mixture, thus,
\beq
  \text{Tr}(\rho^2) = \frac{1}{4}(\text{Tr}(\rho_1^2) +
  \text{Tr}(\rho_2^2) + 2\text{Tr}(\rho_1 \rho_2)).
\eeq
As before, the two lobes are pairwise orthogonal, $\text{Tr}(\rho_1^2) = \text{Tr}(\rho_2^2)$,
and the cross term is zero.  Therefore, we need
\beq
 L= 1 - \frac{1}{2}\text{Tr}(\rho_1^2).
\eeq
The explicit expression for this is the same as in the normal phase,
but with the above factor $1/2$, and the appropriate SR parameters.  In the
large coupling limit, $\text{Tr}(\rho_1^2)=1$,
and thus the linear entropy tends to a constant
$1/2$.  This function, and the finite numerics, are shown in one of the
insets of Fig. \ref{thermvne}.

\subsubsection{Inverse Participation Ratio}
There is also a connection between the linear entropy, and the
inverse participation ratio \cite{Edwards72}, a measure of the
delocalization of a wave function. The un-normalised inverse
participation ratio is defined as, \beq P^{-1} = \int dx dy
|\Psi^4(x,y)|, \eeq Typically, this is normalised over the volume
of the co-ordinate space, however we work with the un-normalised
value for convenience.  $P^{-1}\rightarrow 0$ for a state delocalized across
the entire co-ordinate space, and $P^{-1}$ remains finite for a
localised state, depending on the basis chosen.

We can interpret the participation ratio as a measure of the
spread of a wave function over a particular basis, akin to the way
the entropy is a measure of the spread of a density matrix over
its diagonal basis.  In the normal phase, we perform the Gaussian
integrals with respect to the spin-boson co-ordinates to obtain
\beq \label{part}
  P^{-1} =\frac{\sqrt{\epsilon_- \epsilon_+}}{2\pi}.
\eeq Thus, in this representation, the
participation ratio is equal to the Gaussian normalisation factor of the ground state,
telling us the relative volume in co-ordinate space the state occupies.

 In the SR phase,
$\Psi^{\text{SR}}_G|_{N\gg 1}=\frac{1}{\sqrt{2}}(\psi_1 + \psi_2)$, where
$\psi_{1,2}$ represents the two possible displaced lobes. Again,
using the fact that there is no overlap between these lobes, we
see, \beq P^{-1} = \int dx dy \frac{1}{4}(\psi_{1}^4 + \psi_{2}^4)
= \int dx dy \frac{1}{2}\psi_1^4(x,y), \eeq thus yet again we can
use the normal phase result, with the SR phase parameters.  This
analytical result is plotted in Fig. \ref{thermvne}, where the
delocalization at the critical point is clearly shown.

\begin{figure}[]
\includegraphics[width=0.5\textwidth]{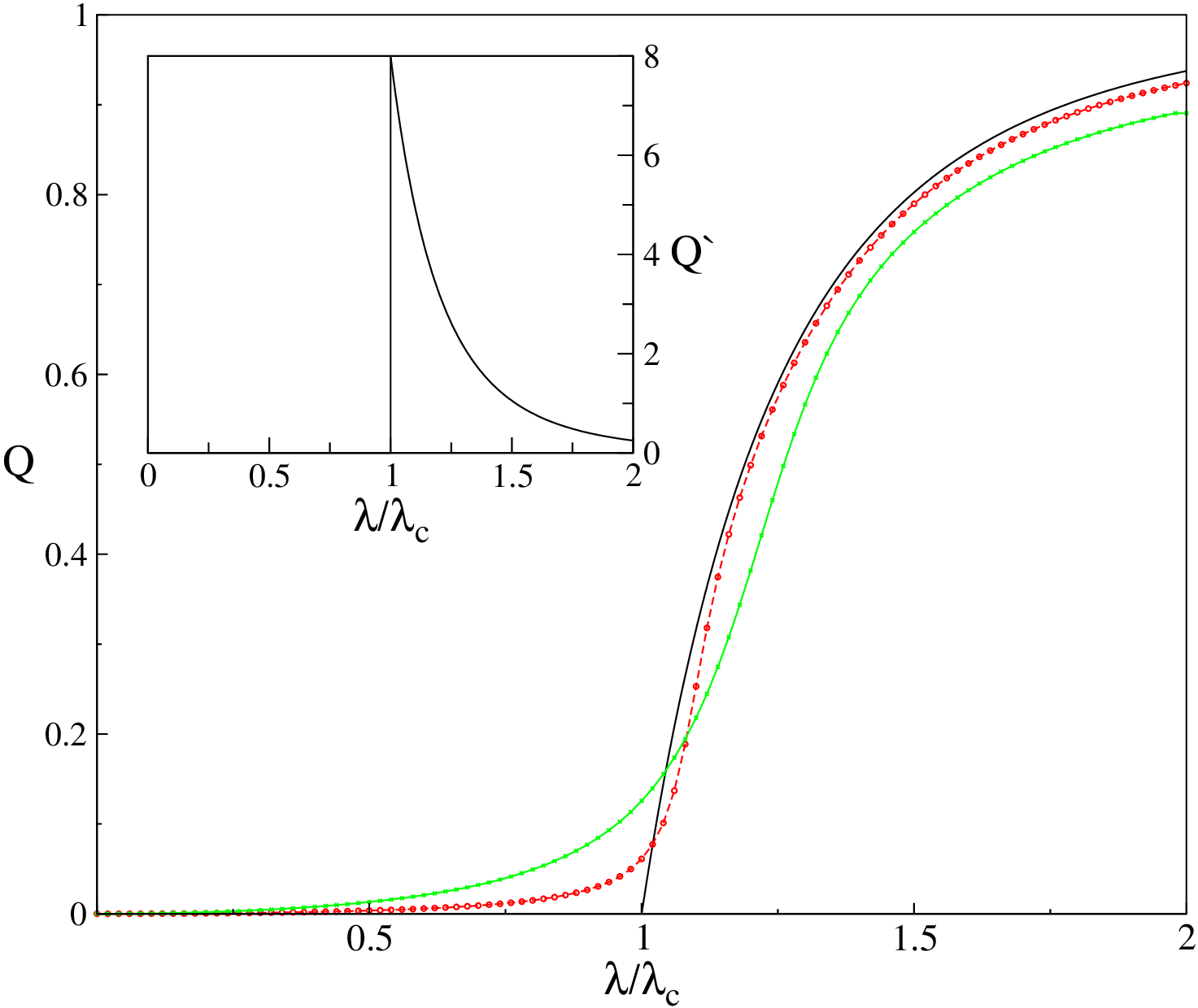}
\caption{The average linear entropy for system sizes $N=8$,
$N=16$, and $N\rightarrow \infty$  Eq. (\ref{qfull}) which includes contributions
from both the field and atomic modes.  Inset:  The derivative of the $N\rightarrow \infty$ limit.} \label{qsym}
\end{figure}

\subsubsection{Average Linear Entropy}

To conclude our discussion, we define the average linear entropy over \textit{all} subsystems ($N$--atoms and the field mode) as,
\beq\label{qfull}
  Q &\equiv& \left[\frac{1}{N+1}[\sum_{k=0}^{N-1}L_k + L_b]\right]= \frac{N}{N+1}L_k + \frac{1}{N+1}(L_b),\nonumber
\eeq
where the sum is replaced because the Dicke states are symmetric with respect to interchange of atoms.  $L_k$ is the linear entropy of atom $k$, and $L_b$ is the linear entropy of the mode discussed earlier,
\beq
        L_k&=& \eta_2 (1-\text{Tr}(\rho_k^2)), \quad L_b=\eta_{N+1}(1-\text{Tr}(\rho_b^2)).
\eeq
As before in Eq.(\ref{line}), the quantities $\eta_2=2$ and $\eta_{N+1}=1+(1/N)$ provide the correct normalisation. 
Since the linear entropy of one subsystem of a pure state is an entanglement monotone \cite{Emary04} (it does not increase under local operations and classical communication), and a valid entanglement measure, a concave function of the linear entropy is also an entanglement monotone.   The average linear entropy $Q$, as we have defined it here, is a concave function, and is thus itself an entanglement monotone.

We can express $\rho_k$, the reduced density matrix of any atom from the ensemble, in terms
of the collective expectation values,
\beq
\rho_k= \left[
    \begin{array}{cc}
    \frac{1}{2}(1-\frac{2\ex{J_z}}{N}) & \frac{\ex{J_-}}{N} \\
    \frac{\ex{J_+}}{N} & \frac{1}{2}(1+\frac{2\ex{J_z}}{N})
    \end{array}
\right],
\eeq
and we find $\text{Tr}(\rho_k^2)=\frac{1}{2} + \frac{2\ex{J_z}^2}{N^2} + \frac{2\ex{J_-}\ex{J_+}}{N^2}$.
Thus, the linear entropy of a single atom is,
\beq
  \label{qsymeq}
  L_k =1-4\ex{J_z}^2/N^2.
\eeq 
In the thermodynamic limit,
$\ex{J_z}^2=\ex{b^{\dagger}b}^2 -N\ex{b^{\dagger}b} + N^2/4$, and
we have $L_{k,N\rightarrow
\infty}^{\lambda < \lambda_c}= 0$. In the super-radiant phase,
$\ex{J_z}^2 = \ex{d^{\dagger}d}^2 - N\mu \ex{d^{\dagger}d} +
N^2\mu^2/4$, and thus $L_{k,N\rightarrow\infty}^{\lambda >
\lambda_c}=(1-\mu^2)$, ($\mu=\lambda_c^2/\lambda$). 
In both phases the contribution from the mode becomes negligible,
and $Q_{N\rightarrow \infty}=L_k$.  
Numerical results for this quantity are plotted in Fig. (\ref{qsym}).  

It has been shown elsewhere~\cite{Brennen03} that the average linear entropy $Q$ is related to 
a measure of multipartite entanglement proposed by Meyer and Wallach \cite{Meyer01}. 
The concept of multipartite entanglement is a difficult and open one
(e.g.\cite{lin99,Cart99,miy03,ver03}), and while the average linear entropy
is limited in the multipartite states it can classify \cite{Emary04}
it is both easy to
calculate and has proved useful in a variety of contexts contexts
\cite{sco04,Brennen03,Scott03,Somma04}.  For completeness, we provide a definition
of Meyer and Wallach's measure in the appendix.


In our results Fig.(\ref{qsym}), we see a clear discontinuity in $Q$ between the two phases. This follows directly from the
discontinuity of the atomic inversion at the critical point
\cite{Emary02}, and the simple nature of our symmetric atomic states.   As a final point, the derivative of $Q$ in the
SR phase TD limit is $\frac{\partial Q}{\partial \lambda} =
4\lambda_c^4/\lambda^5$, which we plot as an inset in Fig. (\ref{qsym}). In the thermodynamic limit the average entanglement of all
the physical subsystems vanishes in the normal phase, but becomes non-zero in the 
superradiant phase.  However, the maximum of $Q$ is not at the critical point, contrary to the 
bipartite partitions and the pairwise partitions \cite{Lambert04}.



\section{Discussion and Conclusions}\label{secDISC}


Table \ref{table1} presents the most important results for the
entanglement measures we have calculated here (and the concurrence discussed in \cite{Lambert04}), and where
appropriate their derivatives.  In particular, we point out the
importance of the divergences at the critical point, and the
finite size scaling exponents (not discussed here) we calculated in \cite{Lambert04}, and which have
been recently confirmed in \cite{Reslen04}.

\begin{table}[h]
\begin{tabular}{|c|c|c|}\hline
 $f(\lambda)$&$f(\lambda\rightarrow\lambda_c)$&
$N$ Scaling \\
\hline $\epsilon_- $&$ |\lambda_c -\lambda|^{1/2}$ & - \\
$l_-$&$|\lambda_c -\lambda|^{-1/4}$& - \\
S & $-\frac{1}{4}\log_2|\lambda-\lambda_c|$ & $\log_2 N^{(0.14 \pm 0.01)}$ \\
$C_R$ & $1-\frac{\sqrt{2}}{2}$& $N^{-0.25 \pm 0.01}$ \\
$\frac{\partial C_R}{\partial \lambda}$& $|\lambda_c-\lambda|^{-\frac{1}{2}}$&-\\
\hline
\end{tabular}
\caption{Entropy $S$, Eq. (\ref{entropy}), concurrence $C_R$ (from \cite{Lambert04}),  and the average linear entropy Eq. (\ref{qsymeq}), in the Dicke model near the critical point $\lambda \rightarrow \lambda_c$. }\label{table1}
\end{table}

For the atom-boson partition, we calculated the entropy exactly
in the thermodynamic limit and
numerically for finite $N$.  The entropy has a
divergence around $\lambda=\lambda_c$, which follows the
power law divergence of the correlation length $l_- \propto
|\lambda-\lambda_c|^{-1/4}$.  

There is also a correspondence between the divergence of the
spin-boson entanglement (entropy) and the delocalization of the
wave function.  This is highlighted by our results for the
behaviour of the participation ratio, which shows that the ground
state of the Dicke model undergoes a massive delocalization at the
critical point.  Since delocalization is a common property of wave
functions in a quantum chaotic system, our results help strengthen
the understanding of the relationship between entanglement and the
underlying integrable to chaotic transition present in the Dicke
Hamiltonian \cite{Emary02}. However, most \textit{generic}
features of
this relationship are still unknown. The future of this field lies
in closer examination of the underlying semi-classical behaviour in
quantum systems, such as supercritical pitchfork
bifurcations~\cite{Hines03} in co-ordinate space, or phase space.

Similarly, we calculated the average linear entropy, and example of the Meyer-Wallach multipartite
entanglement.  Like our other measures, this displays a clear discontinuity at the critical point.
There is obvious future research to be done in applying the many larger classes
of multipartite entanglement measures \cite{Emary04} in a compact way, and perhaps gaining a clearer understanding of the behaviour
of multipartite entanglement. 
Other avenues of future research may arise from investigating quantum
phase transitions in other spin-boson models
\cite{Emary02,Hines03}. In particular, while the Dicke model has a
natural feature that allows infinite system sizes to be
investigated, there is no reason other spin-boson models with
non-commuting energy and interaction terms which do not have this
integrable limit should not exhibit similar critical entanglement,
with chaotic transitions and level statistics.

\begin{acknowledgments}
This work was supported by projects EPSRC GR44690/01, DFG
Br1528/4-1, the WE Heraeus foundation, and the Dutch Science
Foundation NWO/FOM.
\end{acknowledgments}

\appendix

\section{Meyer-Wallach Entanglement} \label{A}
Here we include the definition of the Meyer-Wallach entanglement measure \cite{Meyer01}, as originally intended for a system of $N$--qubits,
and show its connection to the average linear entropy. 

We can write the pure state of $N$ qubits as $\ket{\psi}=\sum_{b_1,...,b_N}a_{b_1,...,b_N}\ket{b_1,...,b_N}$.  Meyer and Wallach defined two new unnormalised states $\ket{\tilde{u}^k}$ and $\ket{\tilde{v}^k}$ as vectors in $C^{2N-1}$ which are obtained by projecting the original state $\ket{\psi}$ onto the two possible subspaces spanned by the two possible states of the $k-\text{th}$ qubit,
\beq
\ket{\psi} = \ket{0}_k\otimes\ket{\tilde{u}^k} + \ket{1}_k\otimes\ket{\tilde{v}^k}.
\eeq
In the Schmidt decomposition, these two subspaces are orthogonal $\langle \tilde{u}^k|\tilde{v}^k\rangle=0$.  $Q$ itself is defined as,
\beq
Q(\ket{\psi})=\frac{4}{N}\sum_{k=1}^N D(\ket{\tilde{u}^k},\ket{\tilde{v}^k})
\eeq
where $D(\ket{\tilde{u}^k},\ket{\tilde{v}^k})=\sum_{i\leq j}|\tilde{u}^k_i\tilde{v}^k_j - \tilde{u}^k_j\tilde{v}^k_i|^2$ is the generalised
wedge product. 

Brennen proved that each term $D$ in the sum of $Q$ was equal to the linear entropy of the $k$th qubit, and thus $Q(\ket{\psi})$ is equivalent to the average
linear entropy of all the qubits,
\beq \label{defQ}
Q(\ket{\psi}) = 2\left[1-\frac{1}{N}\sum_{k=0}^{N-1}\text{Tr}(\rho_k^2)\right],
\eeq
where $\rho_k$ is the reduced density matrix of the $k$th qubit.  Thus, $Q$ has 
the required properties of an entanglement measure $0\leq Q(\ket{\psi})\leq 1$, $Q(\ket{\psi})=0$ for
product states, $Q(\ket{\psi})=1$ for the reduced density matrix of every qubit being maximally mixed, and 
$Q(\ket{\psi})$ is invariant under local unitaries, both because $D$ is invariant, and $\text{Tr}(\rho_k^2)$ is invariant.

Many pure states fulfil the requirement for maximal qubit mixing, and thus give a value of $Q=1$.  
For example, the 3-qubit GHZ state $(\ket{000} +
\ket{111})/\sqrt{2}$ gives $Q=1$, while the Werner state gives $Q((\ket{100} + \ket{010}
+ \ket{001})/\sqrt{3})=\frac{8}{9}$.

\end{document}